\documentclass[twocolumn]{aastex7}

\usepackage{amsmath}
\usepackage{threeparttable}
\usepackage{multirow}
\usepackage{booktabs} % 用于绘制美观的横线
\usepackage{array}    % 用于控制表格列格式

\begin{document}

\title{Precise parameter determination of the open cluster NGC\,1647 via asteroseismology of $p$-mode pulsators}

\author[0000-0002-0040-8351]{Mingfeng Qin}
\affiliation{Institute for Frontiers in Astronomy and Astrophysics, Beijing Normal University, Beijing 102206, P.~R.~China}
\affiliation{School of Physics and Astronomy, Beijing Normal University, Beijing 100875, P.~R.~China}
\email{202331160013@mail.bnu.edu.com}

\author[0000-0001-8241-1740]{Jian-Ning Fu}
\affiliation{Institute for Frontiers in Astronomy and Astrophysics, Beijing Normal University, Beijing 102206, P.~R.~China}
\affiliation{School of Physics and Astronomy, Beijing Normal University, Beijing 100875, P.~R.~China}
\affiliation{Xinjiang Astronomical Observatory, Chinese Academy of Sciences, Urumqi 830011, Xinjiang, P.~R.~China}
\email{jnfu@bnu.edu.cn}

\author[0000-0002-7660-9803]{Weikai Zong}
\affiliation{Institute for Frontiers in Astronomy and Astrophysics, Beijing Normal University, Beijing 102206, P.~R.~China}
\affiliation{School of Physics and Astronomy, Beijing Normal University, Beijing 100875, P.~R.~China}
\email{weikai.zong@bnu.edu.cn}

\author[0000-0003-3816-7335]{Tianqi Cang}
\affiliation{Institute for Frontiers in Astronomy and Astrophysics, Beijing Normal University, Beijing 102206, P.~R.~China}
\affiliation{School of Physics and Astronomy, Beijing Normal University, Beijing 100875, P.~R.~China}
\email{cangtq@mail.bnu.edu.cn}

\author[0000-0002-0474-0896]{Antonio Frasca}
\affiliation{INAF -- Osservatorio Astrofisico di Catania, Via S. Sofia 78, I-95123 Catania, Italy}
\email{antonio.frasca@inaf.it}

\author{Gang Meng}
\affiliation{Institute for Frontiers in Astronomy and Astrophysics, Beijing Normal University, Beijing 102206, P.~R.~China}
\affiliation{School of Physics and Astronomy, Beijing Normal University, Beijing 100875, P.~R.~China}
\email{981214502@qq.com}

\author[0009-0000-4210-1270,gname='Xiran',sname='Xie']{Xiran Xie}
\affiliation{School of Physics and Astronomy, Beijing Normal University, Beijing 100875, P.~R.~China}
\email{xxie@mail.bnu.edu.cn}

\correspondingauthor{Jian-Ning Fu}


\begin{abstract}

% Asteroseismic analysis of pulsating stars provides a robust physical constraint on cluster ages by probing internal structures to determine the core hydrogen abundance.

Asteroseismology of member pulsators provides a robust physical constraint on cluster parameters by linking internal stellar structures to the global properties of the host cluster. However, the parameters of NGC\,1647 remains poorly constrained due to limited investigation, a situation that cluster asteroseismology can significantly refine. In this study, we identified 271 high-confidential cluster members in NGC\,1647, using HDBSCAN clustering with radial-velocity validation. Its initial age is determined in the range of $\sim$125--280~Myr, derived from isochrone fitting based on multi-survey metallicities (LAMOST, APOGEE, and NOT) and extinction-corrected Gaia photometry. Among the members, we found 96 periodic variables from \textit{TESS} and \textit{K2} photometry, including nine $p$-mode pulsators (five $\delta$~Sct and four hybrid $\delta$~Sct--$\gamma$~Dor stars). Assuming a common cluster age and initial chemical composition, joint asteroseismic modeling is performed based on measured large frequency separations and individual mode frequencies. This yields a metallicity of $\mathrm{[Fe/H]} = -0.08^{+0.04}_{-0.01}$, well consistent with the spectroscopic determinations, and a seismic age of $178^{+11}_{-9}$~Myr, more precise than isochrone-based estimates. This work shows the diagnostic potential of $\delta$~Sct asteroseismology in young open clusters and establishes a high-precision benchmark for future studies of \object{NGC\,1647} and other open clusters.
\end{abstract}



\section{Introduction} 
Open clusters, as the birthplaces of many stars in the Galaxy, provide ideal astrophysical laboratories for studying stellar formation and evolution \citep{2003Lada}. Their member stars share, to the first order, a common age, distance, and initial chemical composition, thereby providing strong constraints on stellar models and evolutionary theory \citep{2014Buckner}. Numerous studies have taken advantage of these properties to investigate different types of stars in open clusters. For example, \citet{2023Guzik} analyzed blue straggler stars in NGC\,6819 based on space-based photometry; \citet{2025Ru} focused on B-type stars in the open cluster NGC\,6834; and \citet{2025Berry} explored the occurrence rate of $\delta$~Scuti pulsators in NGC\,3532. Conversely, many works have used specific stellar populations within clusters to constrain the global cluster parameters themselves. For instance, $\delta$~Scuti stars have been used to date young open clusters such as Trumpler\,10 and Praesepe \citep{2023Pamos}; rotational sequences of fast rotators have provided an alternative age diagnostic for NGC\,2281 \citep{2023Fritzewski}; and eclipsing binary systems have been employed to determine the age and distance of NGC\,2506 \citep{2025Yakut}.

Asteroseismology has opened a new window for studying stellar clusters \citep[e.g.,][]{2019Mo}. Early ground-based studies, such as those by \citet{1998Arentoft} and \citet{2005Arentoft}, provided initial asteroseismic insights into open clusters. The field has been revolutionized by high-precision, long-baseline time-domain photometry from space missions. \citet{2010Stello} utilized one month of \textit{Kepler} photometry of NGC\,6791 and NGC\,6819 to perform the asteroseismic analysis of open clusters, highlighting the unique potential of \textit{Kepler} \citep{2010Borucki, 2014Howell} for this field.
Since then, intensive studies have focused on different types of pulsating stars to conduct asteroseismic investigations of open clusters. For example, \citet{2023Brogaard} analyzed red giants in NGC\,6866 to derive an asteroseismic age, while \citet{2025Li} constrained the age of NGC\,2516 using $g$-mode pulsators. Widely distributed across open clusters, $\delta$~Scuti stars are  frequently utilized and serve as important probes for young and intermediate-age systems. They have been increasingly incorporated into asteroseismic investigations of various open clusters \citep{2023Pamos, 2023Palakkatharappil}. These studies demonstrate that asteroseismology of $\delta$~Scutis serves as an effective approach for refining the fundamental parameters of clusters, providing a robust check on ages derived from other methods.

% Asteroseismology has opened a new window for studying stellar clusters \citep{2019Mo}, 
% driven by the rapid progress enabled by high-precision, long-baseline time-domain photometric data from the \textit{Kepler} \citep{2010Borucki,2014Howell} and \textit{TESS} \citep{2014Ricker} missions. 
% \citet{2010Stello} utilized one month of \textit{Kepler} photometry of NGC\,6791 and NGC\,6819 to perform the first asteroseismic analysis of open clusters, highlighting the unique potential of \textit{Kepler} for this field. 

%Among the various classes of pulsating stars, $\delta$~Scuti variables are of particular interest. These are intermediate-mass stars (A--F spectral types) located on or near the main sequence, exhibiting pressure-mode ($p$-mode) oscillations driven primarily by the $\kappa$-mechanism operating in the He\,II ionization zone \citep{Breger2000,Aerts2010}. Their short-period pulsations, often involving multiple excited modes, allow us to probe the outer stellar envelopes and constrain internal processes such as convection, rotation, and mixing. When found in clusters, $\delta$~Scuti stars become especially valuable, since their pulsation properties can be interpreted in the context of well-determined cluster ages and metallicities, thereby linking stellar physics with population studies.  

\object{NGC\,1647} is an open cluster that exemplifies the challenges of precise parameters determination using traditional techniques. Situated beyond the Taurus dark cloud complex \citep{2005Zdanavi}, its center is located at an RA of $71.48^\circ$ and a Dec of $19.08^\circ$ (J2000.0), at a distance of approximately 586.7\,pc \citep{2018Cantat-Gaudin}. A notable disparity exists in recent age determinations for \object{NGC\,1647}; estimates around \(\sim \)260 Myr \citep[see, e.g.,][]{2021Dias} diverge significantly from results near 360–390 Myr reported by other independent studies \citep[e.g.,][]{2020Cantat-Gaudin,2022Carrera,2022Tarricq}.
%Recent literature, however, presents conflicting age estimates for this cluster: while \citep{2021Dias} and \citet{2023Long} suggested an age of $\sim$260\,Myr, other studies have reported significantly older values near 360--390\,Myr \citep[e.g.,][]{2020Cantat-Gaudin, 2022Carrera, 2022Tarricq}. 
These substantial discrepancies, likely arising from variations in membership determinations, reddening corrections, and the choice of stellar models, underscore the ongoing uncertainty regarding the age of \object{NGC\,1647}. This ambiguity motivates the necessity for an independent and more precise age determination using asteroseismology, which can also constrain other fundamental parameters (e.g., metallicity) that remain poorly investigated.

In this study, we present a precise determination of the fundamental parameters for \object{NGC\,1647} by applying asteroseismic analysis to its population of $p$-mode pulsators. Combining high-precision photometry from the \textit{K2} and \textit{TESS} with spectroscopic observation data from LAMOST, we utilize the large frequency separations($\Delta\nu$, the characteristic spacing between consecutive radial modes; \citealp{2010Aerts}) and frequencies of the $\delta$~Scuti stars to derive a precise and independent age for this cluster. The structure of this paper is as follows: Section~\ref{sec:membership} details the membership selection process. Section~\ref{sec:age} describes the determination of fundamental cluster parameters, including extinction and metallicity, alongside traditional isochrone fitting. In Section~\ref{sec:typos}, we characterize the periodic variables identified among the cluster members. Section~\ref{sec:asteroseismology} provides a comprehensive account of the asteroseismic modeling and age inference. We present a detailed discussion of our asteroseismic results in Section~\ref{sec:discuss}, focusing on the determination of cluster age and metallicity. Finally, our findings and conclusions are summarized in Section~\ref{sec:summary}.

\section{Membership Determination}\label{sec:membership}

\subsection{Initial Data Determination and Clustering with HDBSCAN}\label{sec:initial_sample}

The robust determination of member stars is fundamental to open cluster research, which ensures a more accurate determination of cluster parameters. The unprecedented high-precision astrometry from Gaia DR3 \citep{2023Gaia} enabled an accurate and comprehensive member determination for \object{NGC\,1647}. We first adopted the central coordinates of RA\,=\,$71.48^\circ$, Dec\,=\,$19.08^\circ$ (J2000.0) and a mean parallax of $1.68$ mas for \object{NGC\,1647}, as documented in \citet{2018Cantat-Gaudin}. Then all stars were selected as potential members if they were within 100~pc centered around the central coordinates above. To exclude low-quality members, we selected sources with a signal-to-noise ratio (SNR) greater than 10 in the parallax and in each of the Gaia photometric bands (G, BP, and RP). In addition, we corrected the zero-point offsets of parallax for all sources, following the procedure described by \citet{2021Lindegren}. These processes yielded an preliminary sample of 40,070 stars, hereafter referred as Sample\,1.
%figure11111111
% \begin{figure*}
%     \centering
%     \includegraphics[width=1.0\linewidth]{figure/combined_distribution_rv.png}
%     \caption{Membership selection and radial velocity distribution for \object{NGC\,1647}. 
%     Panel (a) shows the spatial sky distribution (RA vs. Dec); 
%     (b) presents the diagram of proper motions ($\mu_{\alpha} \cos\delta$ vs. $\mu_{\delta}$); 
%     (c) displays the CMD. 
%     In these three panels, grey points represent all potential sources identified by HDBSCAN, blue dots denote candidate members with membership probabilities $P \geq 0.7$, and red highlight RV outliers. 
%     Panel (d) illustrates the RV distributions obtained from Gaia DR3 (purple), LAMOST LRS (orange), and LAMOST MRS (green). 
%     Solid curves represent Gaussian fits to each dataset, with the corresponding mean ($\mu$) and standard deviation ($\sigma$) provided in the legend. 
%    The shaded regions highlight stars located beyond the $5\sigma$ clipping threshold; these RV outliers are consistently marked in red across panels (a), (b), and (c).
%    }
%     \label{fig:membership_rv}
% \end{figure*}

\begin{figure*}
    \centering
    \includegraphics[width=1.0\linewidth]{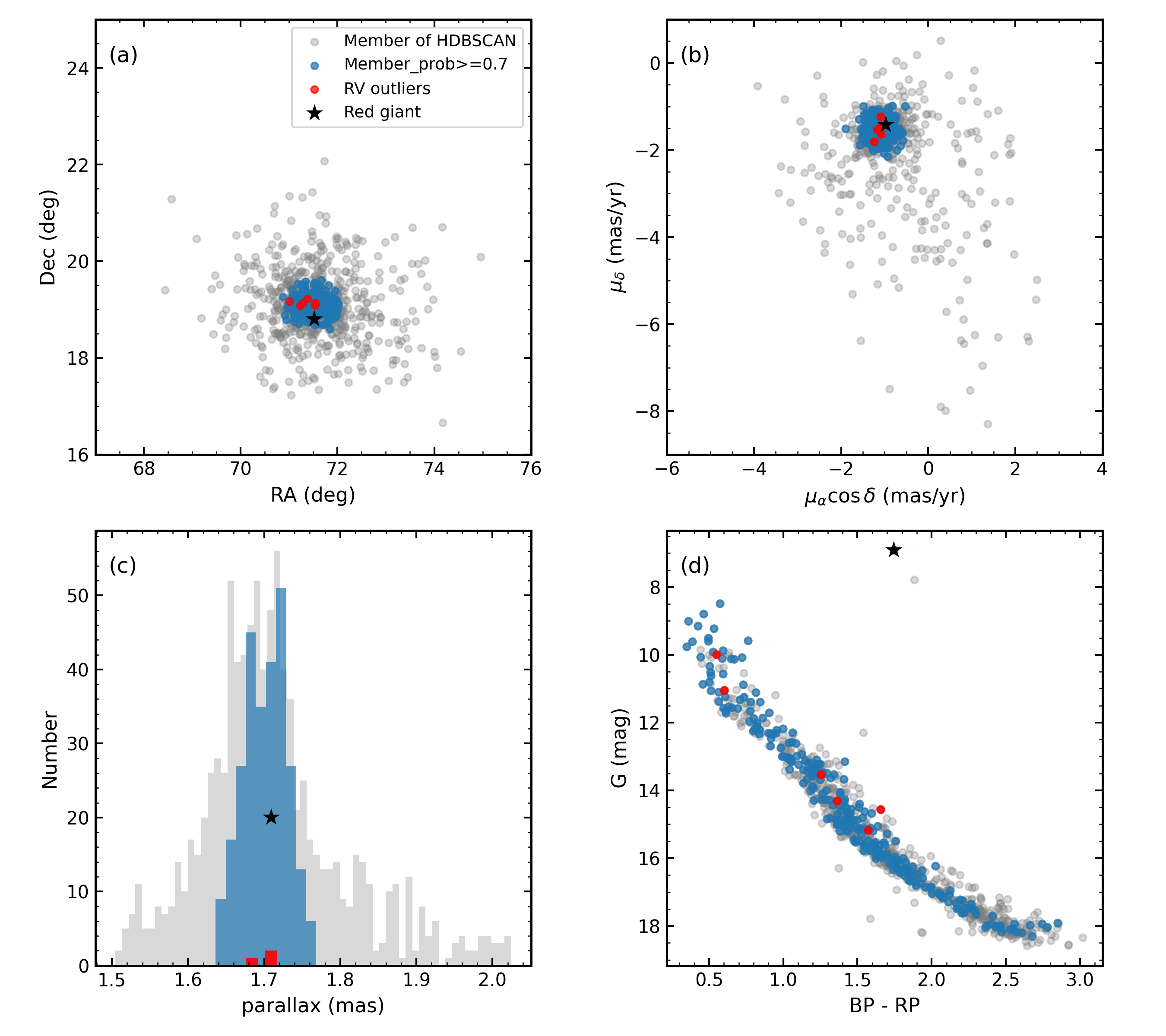}
    \caption{Membership selection for \object{NGC\,1647}.Panel (a) shows the spatial sky distribution (RA vs. Dec), and panel (b) presents the proper motion diagram ($\mu_{\alpha} \cos\delta$ vs. $\mu_{\delta}$); Panel (c) displays the parallax distribution, while panel (d) shows the color--magnitude diagram (CMD, $G$ vs. $G_{BP}-G_{RP}$). In these panels, grey points (or bars) represent all potential sources identified by \texttt{HDBSCAN}, blue denotes candidate members with membership probabilities $P \geq 0.7$, and red highlights RV outliers. The black star denotes the red giant star in NGC 1647.}
    \label{fig:membership}
\end{figure*}

Member stars of an open cluster should originate from the same parent molecular cloud and, consequently, exhibit significant overdensity in both spatial and kinematic spaces. To identify these coherent stellar populations, we employed the \texttt{HDBSCAN} algorithm (\textit{Hierarchical Density-Based Spatial Clustering of Applications with Noise}; \citeauthor{2013Campello} \citeyear{2013Campello}; \citeauthor{2017McInnes} \citeyear{2017McInnes}), which is effective at detecting cluster members within complex and noisy backgrounds. By extending the classical \texttt{DBSCAN} approach, it constructs a cluster hierarchy and extracts the most stable structures based on their persistence across multiple scales. A key advantage of \texttt{HDBSCAN} is its ability to operate without a fixed density threshold, making it robust against irregular stellar distributions often encountered in astrometric data. The algorithm's performance is mainly governed by two parameters: \texttt{min\_cluster\_size}, which sets the minimum threshold for a group of stars to be considered a cluster, and \texttt{min\_samples}, which dictates the algorithm's sensitivity to local noise and outliers.

In our analysis, we applied \texttt{HDBSCAN} to Sample\,1 with a five-dimensional phase space consisting of sky positions ($\alpha, \delta$), parallax ($\varpi$), and proper motions ($\mu_{\alpha^*}, \mu_\delta$). Following the configuration suggested by \citet{2022Tarricq}, we set \texttt{min\_cluster\_size} to 40 and \texttt{min\_samples} to 25. This procedure identified a prominent overdensity corresponding to the spatial and kinematic signature of \object{NGC\,1647}. From the clustering output, we initially extracted a group of 931 stars associated with the primary cluster components, designated as Sample\,2. As shown in Figures~\ref{fig:membership}~(a)--(c), the distribution of Sample\,2 appears relatively extended in both the celestial and kinematic parameter spaces. Furthermore, in the color-magnitude diagram (CMD; Figure~\ref{fig:membership}~(d)), several stars in this sample deviate significantly from the main-sequence (MS) locus. These features suggest that a fraction of field contaminants or low-probability members that the initial clustering could not fully exclude in Sample\,2. To refine the sample, we utilized the membership probabilities ($P$) assigned by \texttt{HDBSCAN}. Following the criteria established in previous studies \citep{2022Tarricq, 2023Hunt}, stars with $P \geq 0.7$ were selected as reliable cluster candidates, which returns a sample of 277 member stars. Compared to the initial clustering output, these candidates (colorful dots) show a significantly more concentrated distribution in all parameter spaces as shown in Figure~\ref{fig:membership}, underscoring their high reliability as intrinsic members of \object{NGC\,1647}.

\subsection{Cleaning and Validation of Candidate Members}
\label{sec:cleaning_validation}
\begin{figure}
    \centering
    \includegraphics[width=\linewidth]{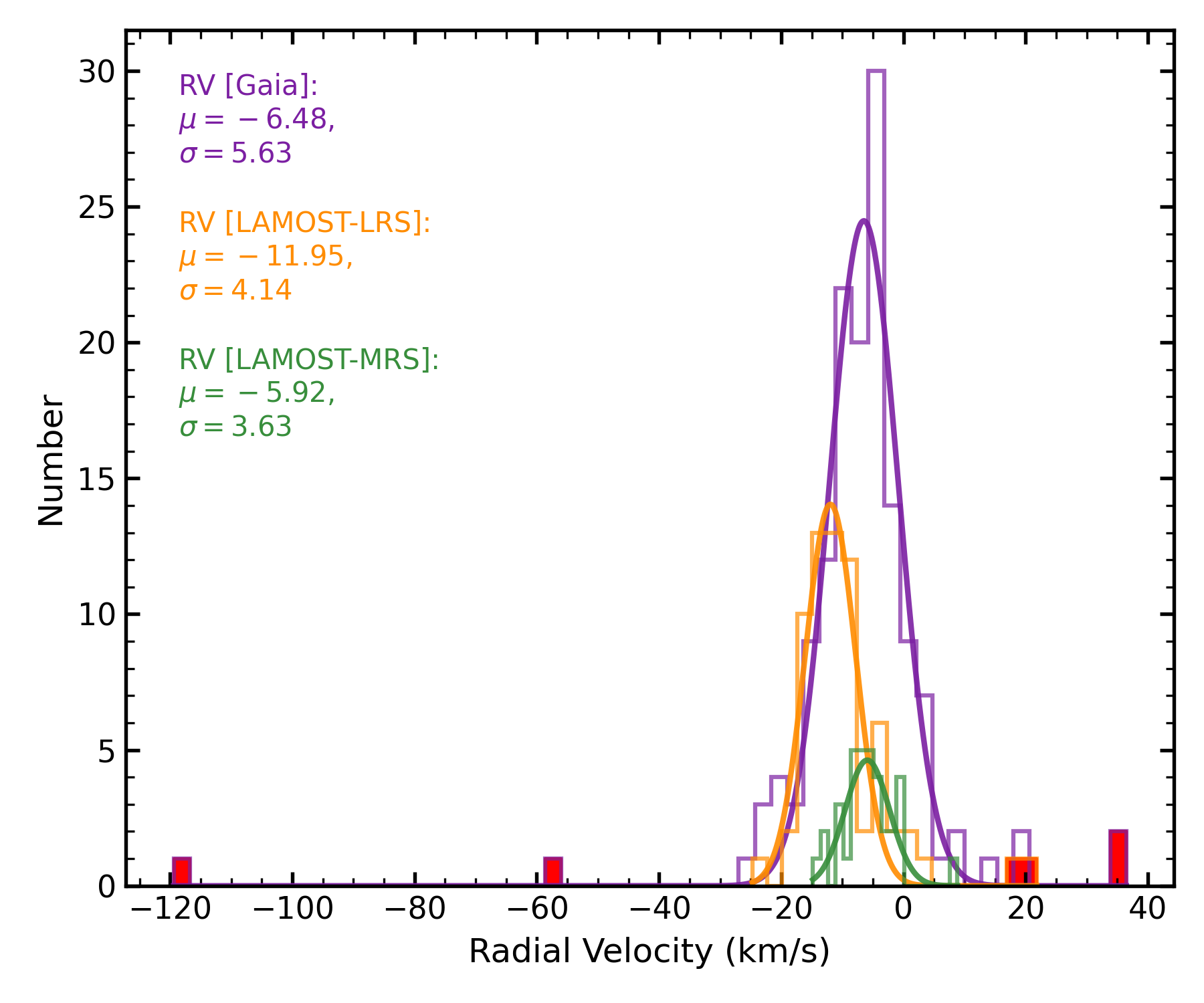}
    \caption{Radial velocity (RV) distributions and Gaussian fitting for \object{NGC\,1647}. 
    % The histograms represent data from \textit{Gaia} DR3 (purple), LAMOST LRS (orange), and LAMOST MRS (green). 
    % Solid curves denote the Gaussian fits to each dataset, with the resulting mean ($\mu$) and standard deviation ($\sigma$) specified in the legend. 
    The shaded red regions highlight stars located beyond the $5\sigma$ clipping threshold, which are identified as RV outliers and excluded from the member list.}
    \label{fig:rv_distribution}
\end{figure}

\setcounter{footnote}{0} % 将计数器重置为 0，下一个脚注就是 1
Although radial velocity (RV) measurements were not included in the initial \texttt{HDBSCAN} clustering due to incomplete data coverage, we utilized available RV information as an independent constraint to further refine our member determination. Theoretically, intrinsic member stars of an open cluster should exhibit consistent RVs reflecting the cluster's bulk motion. To this end, we incorporated RV measurements from three independent sources: \textit{Gaia} DR3, LAMOST DR12 low-resolution spectra \citep[LRS;][]{2012Cui,2012Zhao}, and LAMOST DR12 medium-resolution spectra \citep[MRS;][]{2020Liu}\footnote{\url{http://www.lamost.org/dr12/v1.0/}}. These datasets provided RV information for 144, 66, and 35 stars in our candidate list, respectively. As illustrated in Figure~\ref{fig:rv_distribution}, the RV distribution for each source was fitted with a separate Gaussian function. Visually, a systematic offset of $\sim 5.5$--$6\,\rm{km\,s^{-1}}$ is evident in the LAMOST LRS distribution relative to the \textit{Gaia} DR3 and MRS samples. This shift is consistent with the documented radial velocity zero-point (RVZP) offsets inherent in the LRS pipeline \citep[see details in][]{2020Zong}. To account for this known instrumental characteristic, we applied a $5\sigma$ clipping threshold relative to the fitted mean ($\mu$) of each independent dataset, as indicated by the shaded regions in Figure~\ref{fig:rv_distribution}. By centering the selection window on each survey's specific peak, we ensure that the global zero-point shift does not bias our outlier identification. Consequently, stars falling within the $5\sigma$ range were retained, while others discarded. This process eliminated 4 stars from the \textit{Gaia} DR3 sample and 2 stars from the LAMOST LRS sample (no outliers from the MRS sample). 

% These excluded stars exhibit a clear deviation from the cluster’s primary population, as evidenced in Figures~\ref{fig:membership_rv}a--c. 

Following the removal of these kinematic outliers, we finally established a catalog of 271 high-confidence member stars for \object{NGC\,1647}. The coordinates and membership probabilities of these stars are summarized in Table~\ref{tab:members}.
As shown in Figure~\ref{fig:membership}(d), the CMD of these high-confidence members reveals a remarkably clear and coherent stellar sequence, which serves as a visual testament to the robustness of our membership selection. To further validate our results, we compared our final sample with the census provided by \citet{2023Hunt}, who identified 360 cluster members with probabilities $P \geq 0.7$. We find 258 common stars to theirs, with a substantial overlap, validating the robustness of our identified members.% the reliability of our identified members.
%table111111111111
\begin{table*}
\begin{threeparttable}
\centering
\caption{Member candidates of \object{NGC\,1647}.}\label{tab:members}
\begin{tabular}{ccccccccc}
\hline
Gaia Source & RA(J2000.0) & Dec(J2000.0) & $M_G$ & BP-RP & EPIC& TIC & Variable \\
\hline
3410096353900556544 & 71.04767122 & 19.22136916 & 2.6359 & 0.5807 & 247157739 & 18376371 & gamma dor \\
3410077451747880704 & 70.96937553 & 19.09084764 & 2.4674 & 0.3416 & 247144102 & 18376451 &  \\
3410075527602535424 & 70.94892668 & 19.03029563 & 3.2697 & 0.6476 & 247137766 & 18376493 & gamma dor \\
3410073126717441536 & 71.04618204 & 19.01216284 & 5.6582 & 1.2472 &  --- & 18376504 & --- \\
3409693516031329280 & 70.93836614 & 18.79058415 & 2.9270 & 0.5772 & --- & 18376659 & ---\\
3410126139497593600 & 71.30860440 & 19.51081804 & 8.5904 & 1.9643 & --- & 18437592 &--- \\
3410126246872879232 & 71.25140470 & 19.49799426 & 6.9106 & 1.3821 & --- & 18437618 &--- \\
3410111025508793984 & 71.29381869 & 19.37287249 & 6.4750 & 1.2833 & --- & 18437920 &--- \\
3410111055571988096 & 71.26242450 & 19.36893355 & 2.3703 & 0.3237 & 247173226 & 18437922 &--- \\
3410110338314042368 & 71.22452827 & 19.30452092 & 7.8618 & 1.7904 & --- & 18438091 &--- \\
3410109376241364096 & 71.28719944 & 19.28825630 & 4.8800 & 1.0548 & 247164884 & 18438137 & rotating \\
··· & ··· & ··· & ··· & ··· & ··· & ··· & ···\\
\hline
\end{tabular}
\begin{tablenotes}
\item \textbf{Notes.} Column\,1 lists the Gaia source ID for each target, while Columns\,2 and 3 give the celestial coordinates (RA and Decl.; J2000.0). Columns\,4 and 5 provide the extinction-corrected absolute $G$-band magnitude and the dereddened BP-RP color. Columns\,6 and 7 list the K2 (EPIC) and TESS (TIC) identifiers, respectively. The final column indicates the variability classification of each star. Missing entries are marked with “---”. The complete version of this table will be made available online in a machine-readable format.
\end{tablenotes}
\end{threeparttable}
\end{table*}

\section{Isochrone fitting of NGC\,1647}\label{sec:age}

\subsection{Metallicity Estimation and Extinction Correction}\label{sec:mh}

To estimate the metallicity of \object{NGC\,1647}, we first utilized MRS spectra ($R \approx 7500$) from the LAMOST LK-MRS project \citep{2020Zong,2020Fu}. We identified 25 coadded spectra with $SNR > 20$, a threshold that merits the precision required for robust parameter determination, as established by \citet{2025Qin}. For stars with multiple measurements, we derived a weighted average metallicity, employing weights defined as \(w_{i}=1/\sigma _{i}^{2}\), where \(\sigma _{i}\) represents the uncertainty of each individual measurement. This process resulted in [Fe/H] estimates for 17 member stars. Additionally, one member star has metallicity derived from one APOGEE high-resolution spectrum \citep[$R = 22\,500$;][]{2025SDSS}, while another from high-resolution spectra ($R = 67\,000$) using the Fibre-fed Echelle Spectrograph at the Nordic Optical Telescope \citep[NOT;][]{2022Carrera}. Metallicity measurements were thus available for a total of 19 stars.

\begin{figure}[h!]
    \centering
    \includegraphics[width=\linewidth]{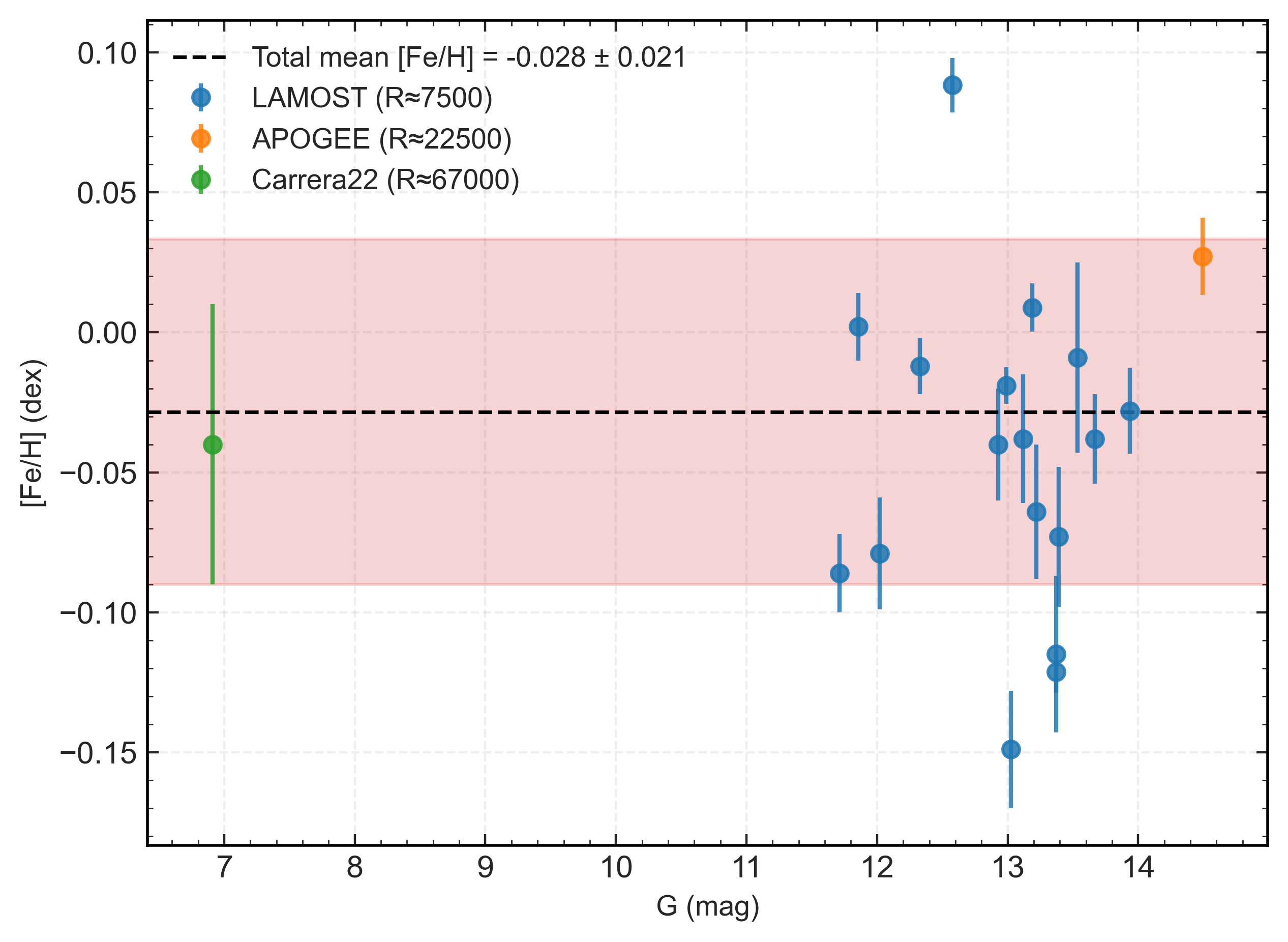}
    \caption{Metallicity as a function of magnitude for stars in \object{NGC\,1647}. Blue dots represent data from the LAMOST MRS, the orange circle denotes the APOGEE measurement, and the green circle corresponds to the data from \cite{2022Carrera}. Error bars indicate the reported uncertainties. The black dashed line marks the weighted mean metallicity, and the pink shaded region represents the $\pm 3\sigma$ confidence interval around the mean.}
    \label{fig:feh}
\end{figure}

We then calculated the average value for the metallicity of \object{NGC\,1647} with associated uncertainty, based on a weighted formalism \citep[see details of Eqn.~1 and 2 in][]{2020Zong}. Accounting for both measurement errors and stellar dispersion, this method provides a representative $\text{[Fe/H]} = -0.028 \pm 0.021$~dex for the cluster. Figure~\ref{fig:feh} displays the mean [Fe/H] value relative to the dispersion of the 19 member stars. The majority of the members lie well within the ±3\(\sigma \) confidence interval, indicated by the shaded region. In contrast, only two outliers fall outside these confidence limits, even when their individual uncertainties are accounted for.

% We then calculated the weighted mean metallicity of the cluster as:
% \begin{equation}
% \langle \mathrm{[Fe/H]} \rangle = \frac{\sum_i w_i \cdot \mathrm{[Fe/H]}_i}{\sum_i w_i},
% \end{equation}
% where $w_i = 1/\sigma_i^2$. To account for the intrinsic spread among the measurements, we computed the weighted sample variance as:
% \begin{equation}
% S_w^2 = \frac{\sum_i w_i (\mathrm{[Fe/H]}_i - \langle \mathrm{[Fe/H]} \rangle)^2}{\left(\sum_i w_i\right)^2 / \sum_i w_i^2 - 1},
% \end{equation}
% and the standard error of the weighted mean as:
% \begin{equation}
% \sigma_{\langle \mathrm{[Fe/H]} \rangle} = \sqrt{\frac{S_w^2}{N_{\mathrm{eff}}}}, \quad \text{where} \quad N_{\mathrm{eff}} = \frac{\left(\sum_i w_i\right)^2}{\sum_i w_i^2}.
% \end{equation}
% This approach ensures that the final uncertainty reflects both the measurement errors and the observed dispersion among individual stars, yielding a robust estimate of the cluster’s mean metallicity.

%figure22222222222222222

% The distribution of stellar metallicities as a function of magnitude is shown in Figure~\ref{fig:feh}. %Blue squares represent data from the LAMOST MRS, the orange circle denotes data from APOGEE, the green circle denotes data from NOT observations, and the error bars indicate the reported uncertainties. 
% The overall weighted mean metallicity is $-0.028 \pm 0.021$. The black dashed line in Figure~\ref{fig:feh} marks this mean value, and the pink shaded region corresponds to the $\pm 3\sigma$ confidence interval.

% \subsection{Extinction Correction}
To account for the interstellar extinction, we adopted a color excess of $E(B-V) = 0.35$, as determined by \citet{2011Guerrero} via multi-band photometry. The extinction within the \textit{Gaia} passbands was computed following the formalism of \citet{2018Danielski}, which relates the monochromatic extinction $A_0$ at 550\,nm to the band-specific extinction (e.g., $A_G$) through the following expression:
\begin{equation}
\begin{aligned}
\frac{A_G}{A_0} = & \, a_1 + a_2 X + a_3 X^2 + a_4 X^3 \\
                  & + a_5 A_0 + a_6 A_0^2 + a_7 A_0^3 \\
                  & + a_8 A_0 X + a_9 X^2 A_0 + a_{10} X A_0^2,
\end{aligned}
\end{equation}
where $X$ represents the intrinsic color index $(G_{\rm BP} - G_{\rm RP})_0$, and the coefficients $a_1-a_{10}$ are specific to the \textit{Gaia} DR3 photometric system \citep{2021Riello}\footnote{\url{https://www.cosmos.esa.int/web/gaia/edr3-extinction-law}}. The absolute magnitudes and intrinsic colors for each member star were subsequently derived by combining individual \textit{Gaia} parallaxes with the calculated extinction values. The resulting de-reddened CMD is presented in Figure~\ref{fig:combined_cmd}~(a). 

% Compared to the observed CMD in Figure~\ref{fig:membership_rv}c, the stellar sequence exhibits a pronounced shift toward bluer colors and higher luminosities, demonstrating the efficiency of our extinction correction in recovering the intrinsic cluster sequence.

\subsection{Isochrone Fitting}
The age of \object{NGC\,1647} was constrained via isochrone fitting based on the MIST stellar evolutionary models \citep{2016Dotter, 2016Choi, 2011Paxton, 2013Paxton, 2015Paxton, 2018Paxton}, with input of $\mathrm{[Fe/H]} = -0.03$~dex. This choice is a compromise between the density of MIST grid with the metallicity resolution of $0.01$~dex and the $\mathrm{[Fe/H]} = -0.028$~dex determined by spectroscopy. 
%While our spectroscopic analysis yielded $\mathrm{[Fe/H]} = -0.028$~dex, . This choice corresponds to the nearest grid point in the MIST isochrones, which are provided at a metallicity resolution of $0.01$~dex.
The model grid spans $\log(\mathrm{Age/yr}) = 5.0$--$10.3$ in steps of 0.05.

To quantify the goodness-of-fit, we employed the squared Euclidean distance, $d^2$, between extinction-corrected magnitude and color index for each star and its nearest neighbor on a theoretical isochrone in the $M_G$ vs. $(G_\mathrm{BP}-G_\mathrm{RP})_0$ plane \citep{2019Liu}. Specifically, to enhance the sensitivity of the age determination and minimize the influence of low-mass stars, the $d^2$ parameter was calculated by restricting the sample to 64 bright stars with $M_G < 3$ mag. The goodness-of-fit parameter is defined as:
\begin{equation}
d^2 = \frac{1}{N} \sum_{i=1}^{N} \left[ \Delta M_{G,i}^2 + \Delta (G_\mathrm{BP}-G_\mathrm{RP})_{0,i}^2 \right],
\end{equation}
where $N$ is the number of stars, and the differences $\Delta$ represent the distance to the nearest neighbor along the isochrone, identified via a $k$-d tree nearest-neighbor search \citep{1975Bentley}.

\begin{figure*}[ht!]
    \centering
    % 请确保文件名与你保存的合并图文件名一致
    \includegraphics[width=1.0\textwidth]{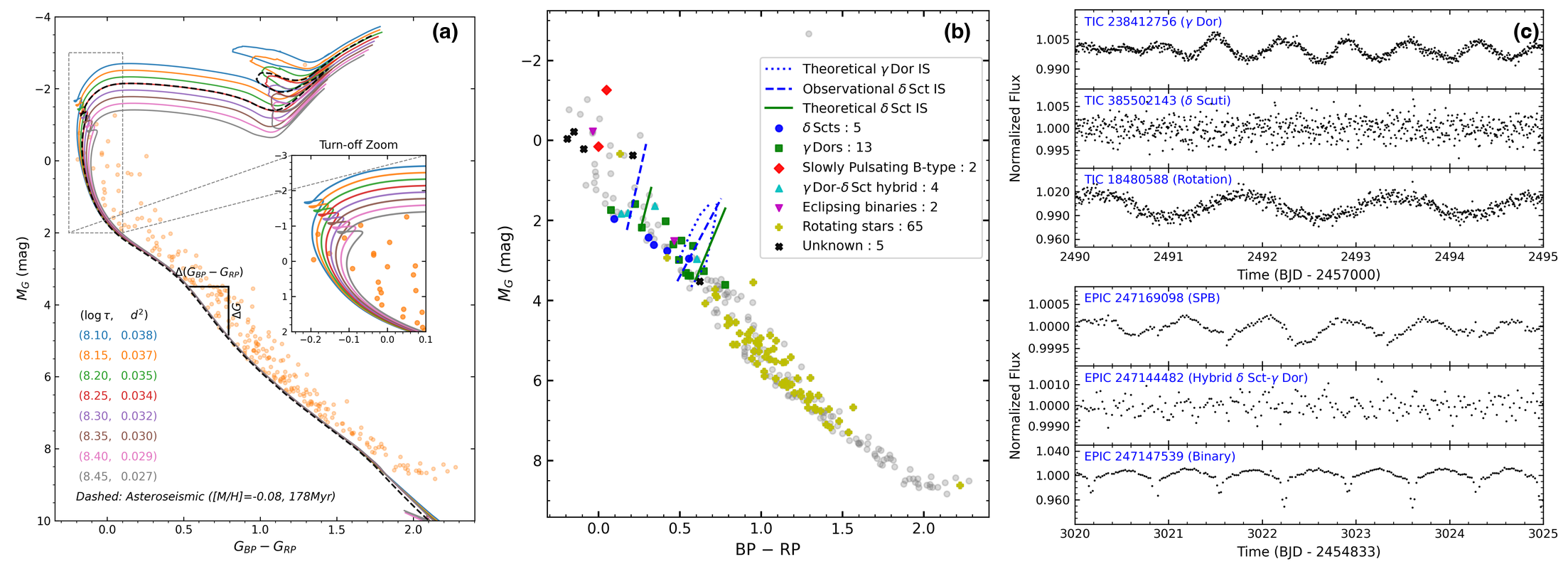}
    \caption{Isochrone fitting and variable stars in \object{NGC\,1647}. 
    (a) Isochrone fitting for \object{NGC\,1647}. Orange points represent extinction-corrected member stars. The solid curves show the eight best-fitting MIST isochrones, where $\tau$ denotes the age of the isochrones.
    (b) CMD of the member stars in \object{NGC\,1647}. Different symbols represent the various types of variable stars identified in this study. The blue dashed and dotted curves indicate the theoretical and empirical instability strips of $\gamma$~Dor and $\delta$~Sct stars, respectively.
    (c) Representative segments of the light curves for the six types of variable stars identified in \object{NGC\,1647}. The top three panels display \textit{TESS} data, while the bottom three panels present observations from the \textit{K2} mission.}
    \label{fig:combined_cmd}
\end{figure*}

Figure~\ref{fig:combined_cmd}~(a) presents the CMD of \object{NGC\,1647}. The cluster exhibits a clear Extended Main Sequence Turn-off (eMSTO) phenomenon at the bright end, and its main sequence (MS) shows a significant width broadening, which we quantify by measuring its vertical and horizontal extents as indicated by the black line segments in the figure, yielding $\Delta G \approx 1.33$ and $\Delta (G_{\rm BP}-G_{\rm RP}) \approx 0.26$. The observed spread likely arises from a combination of photometric uncertainties, differential reddening, and a high fraction of unresolved binaries. To mitigate the systematic bias toward older ages typically induced by these phenomenons, we opted against relying on a single minimum-\(d^{2}\) solution. Instead, we examined a suite of best-fitting isochrones with $d^2 < 0.05$. As shown in the zoom-in panel of Figure~\ref{fig:combined_cmd}~(a), the $d^2$ values for models within the range of $\log(\mathrm{age/yr}) = 8.10$--$8.45$ ($\sim$125--280\,Myr) remain very similar. No significant visual distinction can be made among these isochrones in the turn-off region. This dispersion, likely driven by the eMSTO effect, results in a relatively broad age constraint when relying solely on CMD-based isochrone fitting. Furthermore, the low sensitivity of isochrone morphology to age within the MSTO region, coupled with variations in adopted metallicities and extinction corrections, likely explains the discrepant age estimates reported in previous studies, such as 256\,Myr \citep{2021Dias} and 390\,Myr \citep{2022Carrera}.

%figure4444444444444444444
% \begin{figure}
%     \centering
%     \includegraphics[width=0.48\textwidth]{figure/ngc1647_age.png}
%     \caption{Isochrone fitting for \object{NGC\,1647}. Orange points represent extinction-corrected member stars. The solid curves show the eight best-fitting MIST isochrones, where $\tau$ denotes the age of isochrones.
%     }
%     \label{fig:age}
% \end{figure}

\section{Variable stars in NGC\,1647}\label{sec:typos}

\subsection{Photometric Analysis}
% \footnote{DOI: \url{https://doi.org/10.17909/T9WS3R}}
% \footnote{DOI: \url{https://doi.org/10.17909/t9-nmc8-f686}}

\object{NGC\,1647} is observed by \textit{K2} Campaign~13 with available photometry for 56 member stars, whose light curves  were retrieved from the Mikulski Archive for Space Telescopes (MAST) \citep{K2LightCurves} via the \texttt{Lightkurve} package \citep{2018Lightkurve}. For \textit{TESS}, the available SPOC light curves \citep{2020Caldwell} for 172 member stars were retrieved from MAST \citep{TESSLightCurves}, as they were observed through Sectors\,43-44 and 70-71. All those light curves are processed through the \texttt{Lightkurve} package \citep{2018Lightkurve}, including outlier rejection via a local $3\sigma$ moving clipping algorithm and the \texttt{flatten} routine from the \texttt{wotan} package \citep{Hippke2019} to remove long-term trends. The light curves from different sectors were concatenated and normalized to produce a continuous time series photometry. Further details of these procedures are provided in \citet{2024Xing}. Finally, we computed the Lomb-Scargle periodogram \citep{1976Lomb, 1982Scargle} for each assembled light curve, which were utilized to characterize the variability properties.

To further characterize the variability properties, variable candidates were cross-matched with \textit{Gaia} DR3 to construct a CMD. Their positions on CMD were compared against the theoretical and empirical instability strips for $\delta$~Scuti and $\gamma$~Doradus stars \citep{2005Dupret,2019Murphy}. We note that the instability strip boundaries were transformed from the $T_{\mathrm{eff}}$--$L$ plane into extinction-corrected $(G_{\mathrm{BP}}-G_{\mathrm{RP}})_0$ colors and absolute $M_G$ magnitudes accounting for bolometric corrections and interstellar extinction \citep{2021Riello}. These strips were subsequently superimposed on the CMD for a direct comparison between the observed stellar variability and theoretical predictions.

\subsection{Identification and Classification of Variable Stars}

The variability of each target was characterized by combining the frequency analysis of its light curve with its evolutionary stage on the CMD. We identified 96 periodic variables and assigned them to specific classifications. The sample is dominated by 65 rotational variables, with the remaining fraction comprising various types of pulsators and eclipsing binaries. Figure~\ref{fig:combined_cmd}~(b) presents their distribution on the CMD with with their classifications labelled, while Figure~\ref{fig:combined_cmd}~(c) provides characteristic light curve segments for each class with TESS and K2 photometry. We describe each group of variables in detail as follows.
%The distribution of these variables on the CMD is displayed in Figure~\ref{fig:combined_cmd}~(b), and Figure~\ref{fig:combined_cmd}~(c) illustrates representative light curve segments for each class derived from both \textit{TESS} and \textit{K2} observations.

\medskip\noindent\textbf{Slowly Pulsating B-type (SPB) stars.}
SPB represents a distinct class of early-type variables, exhibiting multiperiodic, high-radial-order \(g\)-mode pulsations. These oscillations provide a window into stellar interior, enabling precise constraints on internal rotation and mixing processes \citep{1991Waelkens}. Two members of \object{NGC\,1647}, TIC\,18442268 and TIC\,18442894, are identified as SPB stars. While the former was previously categorized as a \(\delta \) Sct, \(\gamma \) Dor, or SX Phe candidate \citep{2023Mowlavi}, its location above those instability strips and the presence of low-frequency \(g\)-mode pulsations favor its reclassification as an SPB star. TIC\,18442894, situated near the upper main sequence, is a new identification from this work.

\medskip\noindent\textbf{$\gamma$~Doradus stars.} We discovered 
13 new $\gamma$~Doradus variables as members of \object{NGC\,1647}. Their amplitude spectra exhibit low-frequency $g$-mode pulsations characteristic of $\gamma$~Dor stars \citep{2005Dupret,2015Van}. 
While a few candidates are situated slightly outside the theoretical instability strip, these deviations likely stem from residual uncertainties in interstellar extinction and photometry. Similar discrepant cases have been reported in previous cluster studies, potentially linked to enhanced metallicities and unresolved binary components \citep{2020Li}.
%All identified $\gamma$~Dor variables are confirmed as new members of \object{NGC\,1647}.

\medskip\noindent\textbf{$\delta$~Scuti stars.}
Five new \(\delta \) Scuti stars were identified, exhibiting high-frequency \(p\)-mode pulsations characteristic of this variable class \citep{2019Murphy}. Additionally, three new \(\delta \) Sct--\(\gamma \) Dor hybrid stars were discovered, dominated by high-frequency \(p\)-modes alongside lower-amplitude \(g\)-modes, while an additional hybrid pulsator shows comparable amplitudes in both frequency range. These stars are situated within the overlapping region of the \(\delta \) Sct and \(\gamma \) Dor instability strips, a domain where both the \(\kappa \)-mechanism and convective blocking are effectively operative \citep{2010Grigahc}.

% % figure5555555555555555
% \begin{figure}[ht!]
%     \centering
%     \includegraphics[width=1.0\linewidth]{figure/variable_stars.png}
%     \caption{CMD of the member stars in \object{NGC\,1647}. Different symbols represent the various types of variable stars identified in this study. The blue dashed and dotted curves indicate the theoretical and empirical instability strips of $\gamma$~Dor and $\delta$~Sct stars, respectively.}
%     \label{fig:variable_star}
% \end{figure}

\medskip\noindent\textbf{Eclipsing binaries.}
Two eclipsing binary systems were identified: EPIC\,247147539 and TIC\,18479179. Both are listed in the \textit{Gaia} catalogue of eclipsing binary candidates \citep{2023Mowlavi}, the orbital periods of them is 1.37 and 12.23 day, respectively.

% , this is consist with \citet{2023Mowlavi}. Regarding TIC\,18479179, the light curve exhibits distinct eclipsing events; however, the out-of-eclipse regions are characterized by significant and irregular variability. In contrast, \citet{2023Mowlavi} reported an orbital frequency of 0.05862,d$^{-1}$ for this source, a value that deviates substantially from our primary detected frequency.} 

\medskip\noindent\textbf{Rotational modulation variables.}
Their periodic brightness variations originate from surface inhomogeneities, such as starspots, which rotate in and out of the line of sight \citep{2014McQuillan}. A sample of 65 stars exhibit sinusoidal or quasi-sinusoidal variations, consistent with rotational modulation induced by starspots. The derived rotation periods establish a foundation for gyrochronological dating of NGC 1647, which will be explored in a forthcoming paper.

\medskip\noindent\textbf{Unclassified variables.}
Five stars are identified as variables but defy specific classification with the available photometry.

\section{Asteroseismology of $\delta$ Scuti stars}\label{sec:asteroseismology}

\subsection{Mode identification}

\begin{table*}[htbp]
\centering
\caption{Astrometric and photometric parameters of the 9 target stars in \object{NGC\,1647}.}
\label{tab:target_stars}
\begin{tabular*}{\textwidth}{@{\extracolsep{\fill}}lccccccccc}
\hline
\hline
TIC ID & RA & Dec & Parallax & $\mu_{\alpha}\cos\delta$ & $\mu_{\delta}$ & $G$ & $BP-RP$ & $M_G$\\
 & (deg) & (deg) & (mas) & (mas\,yr$^{-1}$) & (mas\,yr$^{-1}$) & (mag) & (mag) & (mag) \\
\hline
18438742  & 71.2558 & 19.0406 & 1.729 & $-1.084$ & $-1.471$ & 12.611 & 1.085 &2.966 \\
18442420  & 71.3592 & 19.0945 & 1.668 & $-0.895$ & $-1.938$ & 11.394 & 0.844 &1.635\\
18478821  & 71.5460 & 19.3013 & 1.684 & $-1.571$ & $-1.886$ & 12.174 & 0.810 &2.431\\
18479154  & 71.5448 & 18.9735 & 1.719 & $-0.808$ & $-1.741$ & 11.559 & 0.655 &1.835\\
18480650  & 71.6381 & 18.9810 & 1.695 & $-0.940$ & $-1.942$ & 11.722 & 0.614 &1.962\\
18480935  & 71.6072 & 19.2447 & 1.702 & $-1.276$ & $-1.804$ & 12.464 & 0.915 &2.760\\
18540036  & 71.9499 & 19.3833 & 1.680 & $-0.932$ & $-1.287$ & 11.581 & 0.691 &1.814\\
18547844  & 72.0023 & 19.0241 & 1.704 & $-1.062$ & $-1.438$ & 12.327 & 0.839 &2.613\\
385502143 & 71.1643 & 19.2234 & 1.738 & $-1.175$ & $-1.470$ & 12.598 & 1.038 &2.958\\
\hline
\end{tabular*}
\tablecomments{Astrometric and photometric parameters (RA, Dec, Parallax, Proper Motions, $G$ magnitude, and $BP-RP$ color) are retrieved from Gaia DR3. The last column lists the absolute magnitudes obtained in Section~\ref{sec:mh}.}
\end{table*}

Following the classification in Section~\ref{sec:typos}, we identified nine $p$-mode pulsators in \object{NGC\,1647}, comprising five classical $\delta$~Scuti stars and four $\delta$~Sct--$\gamma$~Dor hybrid pulsators. Their fundamental properties, including Gaia DR3 astrometry and photometry, are summarized in Table~\ref{tab:target_stars}.
In \(\delta \) Scuti stars, p-mode frequencies often exceed the Nyquist limit of Kepler long-cadence photometry, giving rise to potential aliasing artifacts known as super-Nyquist frequencies \citep[SNFs;][]{2013Murphy, 2025Wang}. To preclude ambiguities in mode identification and ensure the detection of real pulsation signals, we restrict our analysis to TESS short-cadence photometry.
%Although three of these targets (TIC\,18438742, TIC\,18442420, and TIC\,385502143) were observed by both \textit{TESS} and \textit{K2}, the 1800\,s long-cadence sampling of \textit{K2} results in a Nyquist limit ($\sim 24.5\,\mathrm{d^{-1}}$) that is lower than many of their $p$-mode frequencies. This introduces potential aliasing artifacts, often referred to as super-Nyquist frequencies \citep[SNFs;][]{2013Murphy, 2025Wang}. To avoid such ambiguities in mode identification, we utilized only the \textit{TESS} short-cadence data for these stars.

We performed frequency extraction using the \texttt{MultiModes} package \citep{2022Pamos}, an automated pre-whitening tool specifically designed for the frequency analysis of space-based photometry.
To analyze the extracted frequencies, we searched for regular frequency patterns, such as the distinct ridge structures revealed in \'{e}chelle diagrams. Recent studies have shown that these patterns are frequently observed in young \(\delta \) Sct stars \citep{2020Bedding, 2025Gautam}, offering crucial information to stellar interior. These regularities provide stringent observational constraints necessary for identifying radial ($\ell = 0$) and low-degree non-radial ($\ell=1,2,3$) modes.

\begin{figure}[ht]
    \centering
    \includegraphics[width=0.45\textwidth]{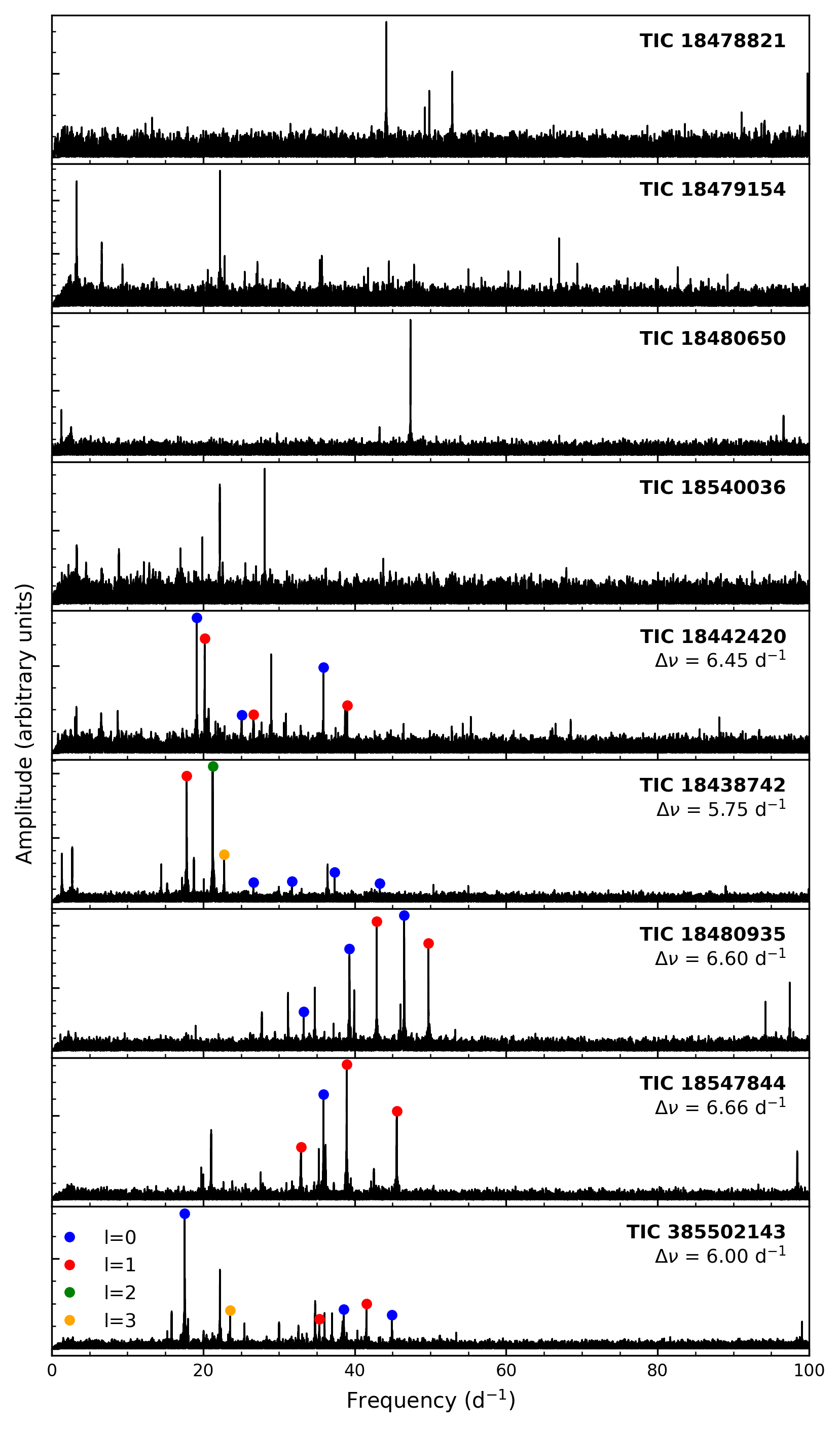}
    \caption{
        Amplitude spectra of the nine $p$-mode pulsators identified in \object{NGC\,1647} based on TESS light curves. The colored symbols denote pulsation modes with different spherical degrees: blue, red, green, and orange circles represent frequencies identified as $\ell = 0, 1, 2,$ and $3$ modes, respectively. For stars exhibiting regular frequency patterns, the measured $\Delta\nu$ values are indicated in the corresponding panels.
    }
    \label{fig:delta_scuti_frequency}
\end{figure}

Then \'{e}chelle diagrams of these nine targets were constructed with the \texttt{\'{e}chelle} Python package \citep{2020Hey}. We have successfully identified five stars exhibiting regular frequency spacings, which we interpret as the large frequency separation (\(\Delta \nu \)); these targets include TIC 18442420, TIC\,18438742, TIC\,18480935, TIC\,18547844, and TIC\,385502143. Using these $\Delta\nu$ values, we performed mode identification for each star, with the identified $\ell = 0$ and $\ell = 1$ frequencies. After this initial observational identification, we further refined the mode identification by iteratively comparing the observed frequency spectra with the theoretical models described in Section~\ref{sec:asteroseismology}. In particular, three frequencies in TIC 18438742 and TIC 385502143 were identified as \(\ell=2\) and \(\ell=3\) modes, primarily based on their agreement with the predicted frequency patterns of the best-fitting models. The identified modes, as marked in Figure~\ref{fig:delta_scuti_frequency}, correspond to the dominant peaks in the spectra and serve as the basis for our asteroseismic modeling. The remaining peaks, some of which exhibit considerable amplitude, are attributed to phenomena such as rotational splitting, modes of higher spherical degree, or the intrinsic mode crowding characteristic of $\delta$ Sct stars{\citep{2022Murphy}}.

However, as illustrated in Figure~\ref{fig:delta_scuti_frequency}, only a limited number of frequencies could be definitively identified for each star. Relying on such sparse frequency sets to constrain asteroseismic models may result in large parameter uncertainties for individual targets. Considering these five pulsators are members of NGC~1647 with a common age and initial chemical composition, we can jointly model these stars by utilizing the 28 identified frequencies and their respective $\Delta\nu$ values as collective constraints. This ensemble approach provides more stringent and reliable estimates for the fundamental parameters of the cluster.

\subsection{MESA model grid and joint fitting}

Stellar evolutionary models were constructed using the \texttt{MESA} code \citep{2011Paxton, 2013Paxton, 2015Paxton, 2018Paxton, 2019Paxton}, with corresponding $p$-mode pulsation frequencies calculated via the \texttt{GYRE} package \citep{2013Townsend, 2018Townsend, 2023Sun}. 
A series model grid was established by spanning the primary stellar parameter space: mass \(M\in [1.4,2.5]\,M_{\odot }\) (step: \(0.01\,M_{\odot }\)) and metallicity $[\text{Fe/H}] \in [-0.10, +0.04]$\,dex (step: $0.01$). This \([\text{Fe/H}]\) range, informed by the spectroscopic analysis in Section 3.1, covers the \(3\sigma \) uncertainty interval, as denoted by the pink shaded region in Figure~\ref{fig:feh}. For each evolutionary track, stellar ages, $\tau$, were sampled up to $500$,Myr with a $1$\,Myr resolution—a range that broadly covers the typical cluster ages reported in the literature ($\lesssim 400$\,Myr). Other inputs in \texttt{MESA} remained at their default settings. For each model, the theoretical large frequency separation ($\Delta\nu_{\mathrm{mod}}$) was derived via a linear fit to $\ell = 0$ modes with $n \ge 5$ \citep{2024Panda}; this criterion was extended to $n \ge 4$ in cases of sparse modes.

 % Given that these five pulsators are confirmed members of \object{NGC\,1647}, we assume they share a common age and initial chemical composition, with their primary differences arising from their stellar masses. Consequently, we utilized both the individual observed frequencies and the derived $\Delta\nu$ to jointly constrain the stellar models. This ensemble approach enables us to simultaneously determine the age and metallicity of the cluster with improved precision.

 %the model grid was partitioned into discrete ([Fe/H], $\tau$) pairs. Within each parameter pair, the stellar mass for each star $j$ was treated as a local free parameter. By minimizing the individual $\chi^2$ values, we could simultaneously determine the cluster's fundamental parameters using the combined frequency sets, thereby overcoming the limitations of the limited number of modes identified for individual stars 

The confirmed cluster membership of these five pulsators provides a unique advantage, enabling a joint fitting analysis by leveraging their coeval and chemically homogeneous nature as collective constraints. Within this framework, we explored the parameter space across a discrete grid of ([Fe/H], $\tau$) pairs. At each grid point, the stellar mass ($M$) for the $j$th star was treated as a local free parameter. To assess the model-data consistency, we calculated a $\chi^2$ statistic for the $j$th star with $N_{\mathrm{obs},j}$ identified frequencies ($\nu_{\mathrm{obs},j,i}$), defined as:
\begin{equation}
\begin{aligned}
\chi^{2}_{j}(\text{[Fe/H]}, \tau; M) = \frac{\bigl(\Delta\nu_{\mathrm{obs},j} - \Delta\nu_{\mathrm{mod}}\bigr)^{2}}{\sigma_{\Delta\nu}^{2}} \\
+ \frac{1}{N_{\mathrm{obs},j}} \sum_{i=1}^{N_{\mathrm{obs},j}} \frac{\bigl(\nu_{\mathrm{obs},j,i} - \nu_{\mathrm{mod,closest},i}\bigr)^{2}}{\sigma_{\nu}^{2}},
\end{aligned}
\end{equation}
where $\nu_{\mathrm{mod,closest},i}$ denotes the model frequency that best matches the $i$-th observed frequency for the given stellar parameters. Regarding the uncertainty terms, we adopt a conservative $\sigma_{\Delta\nu} = 0.2\,\mathrm{d^{-1}}$, a value ten times the bin size of our \'{e}chelle analysis. This accounts for the inherent ambiguity in identifying $\Delta\nu$ for these pulsators. For individual frequencies, we set $\sigma_{\nu} \approx 0.0185\,\mathrm{d^{-1}}$, corresponding to the frequency resolution of two-sector \textit{TESS} observations.

For each star, the best-fitting model mass at a given ([Fe/H], $\tau$) is determined by minimizing the individual $\chi^2$:
\begin{equation}
\chi^{2}_{j,\min}(\text{[Fe/H]}, \tau) = \min_{M} \left[ \chi^{2}_{j}(\text{[Fe/H]}, \tau; M) \right].
\end{equation}
The total $\chi^2$ for each ([Fe/H], $\tau$) group is then computed by averaging the individual minima over the five pulsators:
\begin{equation}
\chi^{2}_{\mathrm{tot}}(\text{[Fe/H]}, \tau) = \frac{1}{5} \sum_{j=1}^{5} \chi^{2}_{j,\min}(\text{[Fe/H]}, \tau).
\end{equation}
This procedure is repeated across the entire ([Fe/H], $\tau$) grid to yield a comprehensive $\chi^2_{\mathrm{tot}}$ map, which serves as the basis for the subsequent statistical inference of the cluster's fundamental parameters.

\subsection{Model Fitting Results}\label{sec:results}

\begin{table*}%[t]
\centering
\begin{threeparttable}
\caption{Comparison of observed and model frequencies for $p$-mode pulsators.}
\label{tab:freq_comparison}
\setlength{\tabcolsep}{5pt}
\begin{tabular}{lccccccccc}
\toprule
& \multicolumn{3}{c}{Observation} & \multicolumn{3}{c}{Best-fit Model} & & \\ 
\cmidrule(lr){2-4} \cmidrule(lr){5-7}
TIC & $\nu_{\mathrm{obs}}$ & $\ell$ & $\Delta\nu_{\mathrm{obs}}$ & $\nu_{\mathrm{mod}}$ & $\ell, n$ & $\Delta\nu_{\mathrm{mod}}$ & $|\delta\nu|$ & $M_{\mathrm{model}}$ \\ 
& (d$^{-1}$) & & (d$^{-1}$) & (d$^{-1}$) & & (d$^{-1}$) & (d$^{-1}$) & ($M_{\odot}$) \\ 
\midrule
\multirow{7}{*}{18438742}  & 21.2522 & 2 & \multirow{7}{*}{5.75} & 21.6484 & 2, 1 & \multirow{7}{*}{5.7230} & 0.3962 & \multirow{7}{*}{$2.46_{-0.03}^{+0.01}$}\\
                           & 17.8253 & 1 &                        & 17.7827 & 1, 1 &                         & 0.0426 & \\
                           & 22.7783 & 3 &                        & 22.7045 & 3, 1 &                         & 0.0738 & \\
                           & 37.3628 & 0 &                        & 37.3651 & 0, 5 &                         & 0.0023 & \\
                           & 31.7056 & 0 &                        & 31.9092 & 0, 4 &                         & 0.2036 & \\
                           & 26.6204 & 0 &                        & 26.9066 & 0, 3 &                         & 0.2862 & \\
                           & 43.3006 & 0 &                        & 43.0518 & 0, 6 &                         & 0.2489 & \\ \midrule
\multirow{6}{*}{18442420}  & 19.1388 & 0 & \multirow{6}{*}{6.45} & 19.4816 & 0, 1 & \multirow{6}{*}{6.45} & 0.3429 & \multirow{6}{*}{$2.17_{-0.04}^{+0.01}$} \\
                           & 20.1908 & 1 &                        & 20.0636 & 1, 1 &                         & 0.1271 & \\
                           & 35.8703 & 0 &                        & 35.9580 & 0, 4 &                         & 0.0877 & \\
                           & 38.9969 & 1 &                        & 38.8609 & 1, 4 &                         & 0.1360 & \\
                           & 26.6129 & 1 &                        & 26.3309 & 1, 2 &                         & 0.2820 & \\
                           & 25.0714 & 0 &                        & 25.0954 & 0, 2 &                         & 0.0240 & \\ \midrule
\multirow{5}{*}{18480935}  & 46.5436 & 0 & \multirow{5}{*}{6.60} & 46.3147 & 0, 5 & \multirow{5}{*}{6.88} & 0.2289 & \multirow{5}{*}{$1.89_{-0.05}^{+0.01}$} \\
                           & 42.9043 & 1 &                        & 42.7813 & 1, 4 &                         & 0.1230 & \\
                           & 49.7456 & 1 &                        & 49.7970 & 1, 5 &                         & 0.0515 & \\
                           & 39.3147 & 0 &                        & 39.5794 & 0, 4 &                         & 0.2646 & \\
                           & 33.2759 & 0 &                        & 33.4028 & 0, 3 &                         & 0.1268 & \\ \midrule
\multirow{4}{*}{18547844}  & 38.9402 & 1 & \multirow{4}{*}{6.66} & 39.0264 & 1, 4 & \multirow{4}{*}{6.47} & 0.0862 & \multirow{4}{*}{$2.16_{-0.03}^{+0.01}$} \\
                           & 35.8646 & 0 &                        & 36.1123 & 0, 4 &                         & 0.2477 & \\
                           & 45.5543 & 1 &                        & 45.4339 & 1, 5 &                         & 0.1204 & \\
                           & 32.9036 & 1 &                        & 32.6963 & 1, 3 &                         & 0.2072 & \\ \midrule
\multirow{6}{*}{385502143} & 17.5627 & 0 & \multirow{6}{*}{6.00} & 17.8805 & 0, 1 & \multirow{6}{*}{5.93} & 0.3178 & \multirow{6}{*}{$2.38_{-0.04}^{+0.01}$} \\
                           & 41.5496 & 1 &                        & 41.5596 & 1, 5 &                         & 0.0100 & \\
                           & 38.5756 & 0 &                        & 38.7134 & 0, 5 &                         & 0.1378 & \\
                           & 23.5534 & 3 &                        & 23.4655 & 3, 1 &                         & 0.0878 & \\
                           & 44.9238 & 0 &                        & 44.6056 & 0, 6 &                         & 0.3182 & \\
                           & 35.3245 & 1 &                        & 35.6911 & 1, 4 &                         & 0.3666 & \\ \bottomrule
\end{tabular}
\begin{tablenotes}
      \small
      \item \textbf{Notes.} Observed pulsational properties are compared with the best-fit model predictions. Columns 2–4 list the observed frequencies ($\nu_{\mathrm{obs}}$), identified spherical degrees ($\ell$), and $\Delta\nu_{\mathrm{obs}}$. Columns 5–7 provide the corresponding theoretical values, where $\ell$ and $n$ denote the spherical degree and radial order, respectively. $|\delta\nu|$ represents the absolute frequency residual ($|\nu_{\mathrm{obs}} - \nu_{\mathrm{mod}}|$). The asteroseismic stellar masses ($M_{\mathrm{model}}$) in the final column are derived from the 68\% (1$\sigma$) HPD credible region of the joint fitting solution.
\end{tablenotes}
\end{threeparttable}
\end{table*}

\begin{figure}[ht]
    \centering
    \includegraphics[width=\linewidth]{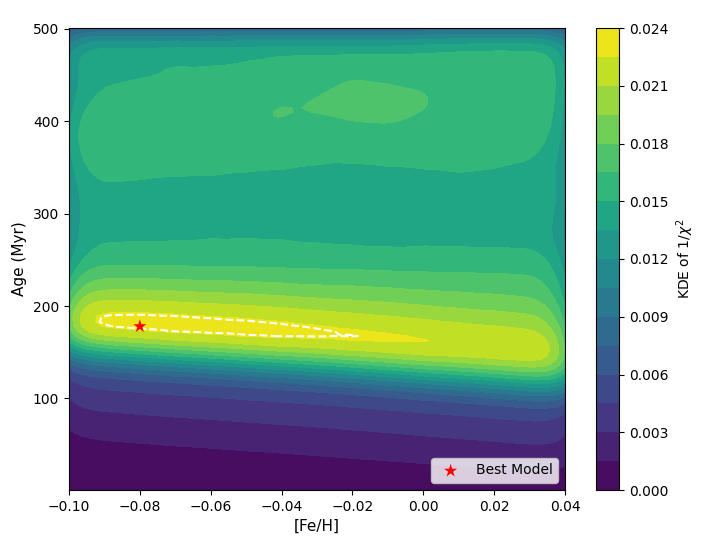}
    \caption{A two-dimensional Kernel Density Estimation (KDE) in the ([Fe/H], $\tau$) parameter space, with each model weighted by its inverse-$\chi^{2}$ value. The white dashed contour marks the $3\sigma$ confidence region defined by $\chi^{2}_{\mathrm{tot}} \le \chi^{2}_{\min} + 9.21$, and the red pentagram indicates the best-fitting model within the grid.
    }
    \label{fig:kde_map}
\end{figure}

\begin{figure*}[ht]
    \centering
    \includegraphics[width=\textwidth]{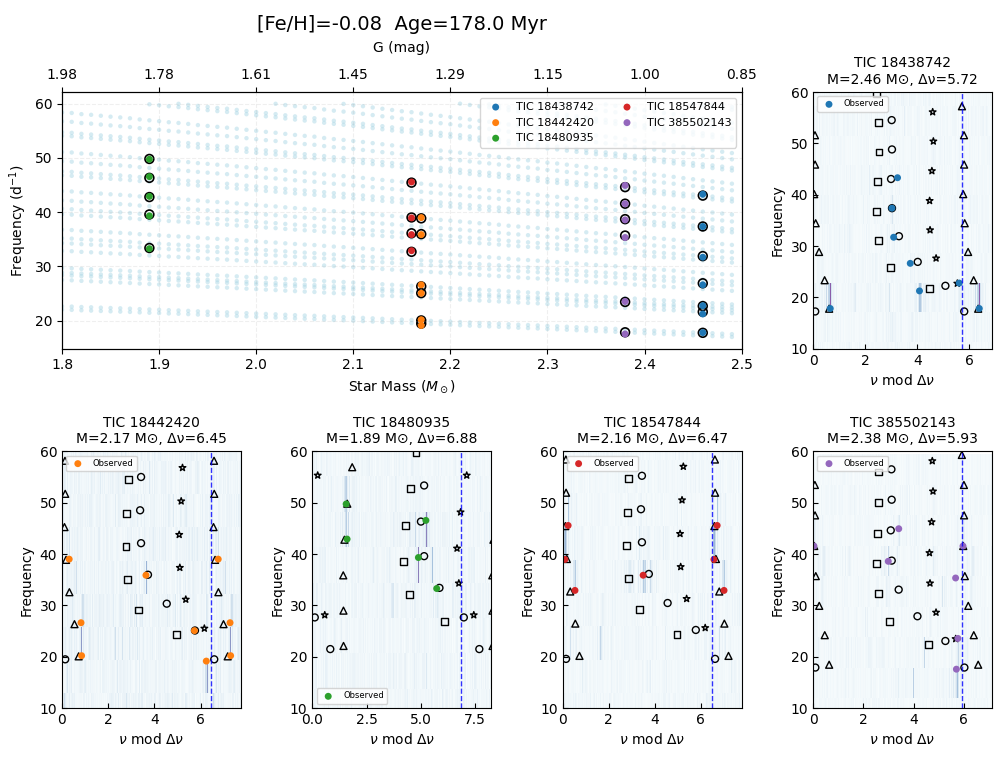}
    \caption{
    Comparison between the observed and modeled pulsation properties for the five pulsating members of \object{NGC\,1647}. 
    Top-right panel: comparison between the observed (filled symbols) and modeled (open symbols) pulsation frequencies for the best-fitting model of each star. The bottom horizontal axis shows the stellar mass, while the top horizontal axis indicates the corresponding Gaia $G$-band absolute magnitude derived from the models. Black open circles mark the model frequencies that best match each observed frequency. 
    Top-left and bottom panels show the individual \'echelle diagrams for the five pulsators, where the observed frequencies (filled circles) and their closest matching model frequencies (open symbols) are plotted modulo $\Delta\nu$ and overlaid on the corresponding power spectra. 
    Different symbols denote modes with different spherical degrees: circles ($l=0$), triangles ($l=1$), squares ($l=2$), and stars ($l=3$).
    }
    \label{fig:mesa_model}
\end{figure*}

We construct a two-dimensional Kernel Density Estimation (KDE) in the ([Fe/H], $\tau$) parameter space, with each model weighted by its inverse-$\chi^{2}$ value. As shown in Figure~\ref{fig:kde_map}, the resulting distribution exhibits a single, prominent density peak, indicating that the pulsational constraints converge toward a unique solution in the ([Fe/H], $\tau$) plane. To determine the joint uncertainties of the cluster parameters [Fe/H] and $\tau$, we employ a likelihood-ratio test \citep{1976Avni,2024Panda}. By optimizing the stellar masses at each fixed ([Fe/H], $\tau$) grid point, we effectively marginalize their influence, allowing for a direct evaluation of the goodness of fit across the parameter plane. Assuming two degrees of freedom, the $3\sigma$ confidence region is defined by the contour $\chi^{2}_{\mathrm{tot}} \le \chi^{2}_{\min} + 9.21$, marked by white dashed contour in Figure~\ref{fig:kde_map}. Within this boundary, 101 models are retained, corresponding to approximately 1.06\% of the lowest-$\chi^2$ models in the grid, which defines the statistical confidence region around the best-fitting solution. We adopt the KDE peak values as the final cluster parameters. The associated uncertainties are derived from the 68\% highest posterior density (HPD) credible interval, calculated within the $3\sigma$ confidence region. This corresponds to 69 models, from which we obtain: $\tau = 178^{+11}_{-9}$ Myr and $[\mathrm{Fe/H}] = -0.08^{+0.04}_{-0.01}$ dex.

The performance of our best-fitting models is illustrated in Figure~\ref{fig:mesa_model}. In the top-right panel, the observed frequencies (colored dots) are plotted against the theoretical predictions (black open circles). To provide a comprehensive view of the stellar parameters, the bottom axis indicates the model-derived stellar masses, while the top axis shows the corresponding theoretical absolute magnitudes in the $G$ band ($M_G$). These $M_G$ values were calculated by passing the model parameters through the synthetic photometry tool provided by \citet{2019Chen} \footnote{\url{https://sec.center/YBC/}}. The observed frequencies agree well with the theoretical predictions, confirming the robustness of the joint fitting procedure.

As summarized in Table~\ref{tab:freq_comparison}, the joint fitting procedure yields a robust asteroseismic solution for all five pulsators. Quantitatively, the individual frequencies ($\nu_{\mathrm{obs}}$) and large frequency separations ($\Delta\nu_{\mathrm{obs}}$) show mean absolute differences of $0.18\,\mathrm{d^{-1}}$ and $0.10\,\mathrm{d^{-1}}$ relative to the model predictions, respectively. These residuals are comparable to the $0.2\,\mathrm{d^{-1}}$ resolution limit of the \'{e}chelle analysis used for mode identification. Such consistency is visually manifested in Figure~\ref{fig:mesa_model}, where the majority of observed frequency peaks align precisely with the theoretical $\ell = 0$ and $\ell = 1$ ridges constructed using the model $\Delta\nu_{\mathrm{mod}}$. By successfully recovering these ridge structures, the models not only confirm the observed spherical degrees $\ell$ but also provide the constraints to assign radial orders $n$ for all identified modes. This agreement yields a self-consistent seismic solution for the pulsating members of \object{NGC\,1647}.

% \textbf{We compared the Gaia absolute magnitudes listed in Table~\ref{tab:target_stars} with the stellar masses derived from the seismic modeling in Table~\ref{tab:freq_comparison}. The comparison shows that the mass ranking does not strictly follow the expected mass–luminosity relation. For example, TIC\,18442420 and TIC\,18547844 have nearly identical model masses ($2.17$ and $2.16\,M_{\odot}$), but their absolute magnitudes differ significantly ($M_G = 1.635$ and $2.613$). This discrepancy is likely caused by differential extinction within NGC~1647. In Section~\ref{sec:mh}, the absolute magnitudes were derived using a single cluster extinction value, whereas the extinction may vary among individual members, leading to different luminosity offsets. This is consistent with the findings of Frasca et al. (2026, submitted), who demonstrated that NGC~1647 exhibits significant differential extinction.} 

A comparison between the Gaia absolute magnitudes in Table~\ref{tab:target_stars} ($M_G$) and the corresponding model-derived absolute magnitudes ($G$ in Figure~\ref{fig:mesa_model}) reveals a noticeable discrepancy. For instance, TIC\,18442420 and TIC\,18547844 possess nearly identical model-predicted magnitudes ($1.337$ and $1.353$ mag, respectively), whereas their observation-based absolute magnitudes differ by approximately 1.0 mag ($M_G = 1.635$ and $2.613$ mag, respectively). We note that the absolute magnitudes in Section~\ref{sec:mh} were derived assuming a single extinction value for the entire cluster. In this context, Frasca et al. (2026, submitted) found that $A_V$ values in NGC\,1647 exhibit a substantial range from 0.3 to 2.1 mag, which ($\Delta A_V \approx 1.8$ mag) can induce an average shift of approximately 1.3 mag in $M_G$ and 0.8 mag in $(G_{BP}-G_{RP})$. This indicates that differential extinction may be a factor contributing to the discrepancy between the Gaia absolute magnitudes and the seismic model predictions.

\section{Discussion}\label{sec:discuss}

Asteroseismology has proven to be a powerful tool for deriving precise fundamental parameters of open clusters \citep[e.g.,][]{2023Brogaard, 2023Pamos, 2023Palakkatharappil, 2025Li}. \object{NGC\,1647} presents a diverse laboratory for such studies, hosting various types of pulsators including SPB stars, $\gamma$~Doradus stars, and $\delta$~Scuti stars. However, current observational data dictate which of these can be effectively utilized as seismic probes. While SPB and $\gamma$~Doradus stars offer unique insights into deep interiors via $g$-mode pulsations, their characteristic long periods require an observational baseline exceeding our current data to achieve the necessary frequency resolution. Furthermore, although a red giant member is identified in Figure~\ref{fig:combined_cmd}~(b), no detectable solar-like oscillations were found in that light curve. Consequently, the short-period $p$-mode pulsations of the $\delta$~Scuti members are clearly resolved, making them the most robust and practical tools for refining the cluster's fundamental parameters at this stage.

The presence of a red giant in such a young cluster warrants further discussion. The membership of this star is well supported by several independent indicators. Its Gaia DR3 astrometric parameters are consistent with those of the cluster members, and the membership probability is 0.88 (see Figure~\ref{fig:membership}). In addition, its radial velocity ($RV = -6.74$ km\,s$^{-1}$) is consistent with the cluster mean value ($-6.48 \pm 5.63$ km\,s$^{-1}$). From an evolutionary perspective, its position on the CMD (Figure~\ref{fig:combined_cmd}~(a)) agrees well with the best-fitting isochrone ($t \approx 178$ Myr). The corresponding model mass at this evolutionary stage is about $4.17\,M_{\odot}$, indicating that a mid-mass star in the cluster could have already evolved off the main sequence and reached the giant phase. Although this evolutionary stage is relatively short-lived, the presence of such a star is still compatible with the expected evolution of intermediate-mass stars in a $\sim178$ Myr cluster. Therefore, although rare, the existence of this red giant does not contradict the derived cluster age.

Our results for $\delta$~Scuti stars underscore the remarkable capability of asteroseismology to refine cluster parameters with high precision. As illustrated in Figure~\ref{fig:combined_cmd}~(a), the isochrone corresponding to our best-fit asteroseismic age ($178_{-9}^{+11}$\,Myr) aligns well with the age interval independently determined via isochrone fitting in Section~\ref{sec:age} (125--280~Myr). Similarly, the derived metallicity of $\mathrm{[Fe/H]} = -0.07_{-0.01}^{+0.04}$ is consistent with the spectroscopic value obtained in Section~\ref{sec:mh}. This dual-parameter agreement supports the robustness of the seismic solution. However, the model comparison shown in Figure~\ref{fig:kde_map} indicates that the pulsation frequencies are more sensitive to stellar age than to metallicity within the explored parameter range. Consequently, the relatively small uncertainty in $\mathrm{[Fe/H]}$ mainly reflects the distribution of acceptable models within the adopted grid. Systematic uncertainties associated with the input physics of stellar models (e.g., opacities, mixing prescriptions, and convection overshooting) may introduce additional uncertainties in the derived cluster parameters that are not fully captured by the $\chi^2$ analysis.

% Despite the overall success of the fit, we acknowledge the presence of minor discrepancies, with a mean absolute frequency residual of approximately $0.15\,\rm{d}^{-1}$ and a maximum value of $0.5\,\rm{d}^{-1}$. These small offsets likely stem from the simplified physical framework of our models, such as the neglected effects of non-adiabaticity \citep{2005Dupret} and the simplified treatment of convective overshooting. 
% Notably, our models are currently based on a non-rotating configuration. While stellar rotation typically introduces frequency splitting and systematic shifts in pulsation ridges \citep{2010Aerts, 2022Murphy}, we found no compelling evidence of significant rotational splitting in our Fourier analysis. Furthermore, as illustrated in Figure~\ref{fig:mesa_model}, the observed \'{e}chelle ridges align closely with the $m=0$ theoretical predictions, showing none of the characteristic tilting or systematic offsets reported in more rapidly rotating stars . This close alignment suggests that rotational effects do not dominate the pulsation spectra of our targets, justifying our use of non-rotating models and ensuring that the derived cluster parameters remain robust. 

Among the possible sources of systematic uncertainties, stellar rotation and convective core overshooting can significantly affect both stellar evolution and pulsation properties. In the present work, we adopted a simplified grid of stellar models that does not include rotational effects or overshooting. For the five $\delta$~Scuti stars analyzed here, only \object{TIC\,18547844} currently has a spectroscopic measurement of projected rotational velocity, with $v\sin i = 85.4$ km\,s$^{-1}$ derived from LAMOST spectra. Several relatively high-amplitude frequencies in this star (e.g., 21.0544~d$^{-1}$) cannot be reliably reproduced by non-rotating models and may be related to rotational effects. Similar unidentified frequencies are also present in \object{TIC\,18438742} and \object{TIC\,385502143}, suggesting that rotation may influence the pulsation spectra of these stars. Rotation can alter oscillation frequencies through rotational splitting and structural modifications caused by centrifugal distortion and rotational mixing \citep{2022Murphy}, while convective core overshooting affects the core size and evolutionary timescale of intermediate-mass stars. Because these effects are not included in the present models, the uncertainties in the derived cluster parameters, particularly the age and metallicity, may be slightly underestimated. Nevertheless, the good agreement between the seismic results and independent constraints from isochrone fitting and spectroscopy indicates that the simplified modeling still provides meaningful constraints on the global properties of the cluster.

\section{Summary}\label{sec:summary}

In this work, we performed a comprehensive investigation of the open cluster \object{NGC\,1647} by integrating datasets from space-based missions (\textit{Gaia}, \textit{K2}, and \textit{TESS}) and ground-based facilities (LAMOST, APOGEE, and NOT). From an initial catalog of stars within a 100\,pc projected radius of the cluster center, we employed the HDBSCAN algorithm to identify a robust sample of cluster members, adopting a membership probability threshold of $P_{\mathrm{mem}} \ge 0.7$. Subsequent refinement using radial velocities effectively eliminated residual contaminants, resulting in a final sample of 271 high-confidence member stars. The metallicity of \object{NGC\,1647} was determined to be $\text{[Fe/H]} = -0.028 \pm 0.021$~dex based on spectroscopic measurements from LAMOST, APOGEE, and \citet{2022Carrera}. Based on the extinction-corrected magnitudes and color indices of the identified members, traditional isochrone fitting yields a relatively broad age range, spanning from 125 to 280\,Myr.

Leveraging the high-cadence \textit{K2} and \textit{TESS} light curves, we identified 96 periodic variables within the cluster. Among these, nine stars exhibiting $p$-mode pulsations (comprising five classical $\delta$~Sct stars and four hybrid $\delta$~Sct--$\gamma$~Dor stars) provided the basis for our subsequent asteroseismic analysis. We determined the $\Delta\nu$ and identified $\ell=0$ and $\ell=1$ pulsation modes using \'echelle diagram diagnostics. By fitting these seismic constraints against those of a grid of \texttt{MESA/GYRE} evolutionary and pulsation models, we derived a joint asteroseismic solution for the cluster: $\tau = 178^{+11}_{-9}$\,Myr and [Fe/H] = $-0.08^{+0.04}_{-0.01}$\,dex.

% Our asteroseismology-determined metallicity is in excellent agreement with the estimate from spectroscopic observations, while the inferred cluster age is younger than several estimates reported in the literature based on isochrone fitting. Previous studies have typically suggested ages of $\sim 390$\,Myr or $\sim 260$\,Myr; such discrepancies often stem from differing assumptions regarding metallicity, extinction, and membership criteria. Under a uniform extinction treatment and spectroscopically supported metallicity, our seismic solution falls within the age range discussed in Section~\ref{sec:age} but offers a significantly narrower uncertainty interval. This improved precision underscores the capability of $\delta$~Sct asteroseismology to break the degeneracies inherent in traditional CMD analysis.

This asteroseismic metallicity agrees wll with spectroscopic results, while the derived age remains consistent with traditional estimates but offers significantly higher precision. As high-precision time-series photometry continues to grow with future missions like \textit{PLATO} \citep{2014Rauer}, the growing census of seismic members will further bolster the robustness of cluster modeling. Such advancements will enable even more stringent constraints on the fundamental parameters of \object{NGC\,1647} and similar systems, deepening our understanding of stellar and cluster evolution.

%% Please use the acknowledgment and contribution environments. This will 
%% be anonomyized when the "anonymous" style option is used. 
\begin{acknowledgments}

We are grateful to the referee for the constructive comments and helpful suggestions, which significantly improved this manuscript. We acknowledge the support from the National Natural Science Foundation of China (NSFC) through the grants 12427804, 12273002, 12503035 and 12541303. This work is supported by the China Manned Space Program with grant no. CMS-CSST-2025-A13, the Tianchi Talent Introduction Plan, and the Central Guidance for Local Science and Technology Development Fund under No. ZYYD2025QY27. M.F. thanks Haozhi Wang, Gang Li, Enrico Corsaro, and Sylvain N. Breton for their valuable and insightful discussions. 
% Guoshoujing Telescope (the Large Sky Area Multi-Object Fiber Spectroscopic Telescope LAMOST) is a National Major Scientific Project built by the Chinese Academy of Sciences. Funding for the project has been provided by the National Development and Reform Commission. LAMOST is operated and managed by the National Astronomical Observatories, Chinese Academy of Sciences. %T.C. and W.Z. acknowledge funding from the support of the NSFC through grant 12273002.  T.C. acknowledges funding from the support of the National Natural Science Foundation of China (NSFC) through grant 12503035. 
A.F. acknowledges funding from the Large Grant INAF-2024 %“Spectral Key features of Young stellar objects: Wind-Accretion LinKs Explored in the infraRed (SKYWALKER)”.
``SKYWALKER''.
\end{acknowledgments}

% \begin{contribution}
%%This section gives authors the space to recognize author contributions. The text inside this environment is NOT counted towards the total word quanta. At a minimum, manuscripts are expected to include this text:

% All authors contributed equally to the Terra Mater collaboration.

%% But authors are expected to provide more specific details, e.g. 
%%
%%SC was responsible for writing and submitting the manuscript.
%%WWM came up with the initial research concept and edited the manuscript.
%%OTS obtained the funding and edited the manuscript.
%%EBF provided the formal analysis and validation. He also edited the manuscript.
%%GEH Supervised the undergraduates, wrote the software and administers the project github and Zenodo repositories.
%%
%% Authors can use the Contributor Role Taxonomy (CRediT) at
%% for ideas on how write a good statement tailored to their needs.

% \end{contribution}

\facilities{Gaia, Kepler, TESS, LAMOST, APOGEE}

\software{\texttt{HDBSCAN} \citep{2013Campello,2017McInnes},
          \texttt{Lightkurve} \citep{2018Lightkurve}, 
          \texttt{\'{e}chelle} \citep{2020Hey}}

\bibliography{sample7}{}
\bibliographystyle{aasjournalv7}



\end{document}